\documentclass[twocolumn,tighten,times]{aastex631}

\usepackage{graphicx}
\usepackage[utf8]{inputenc}
\usepackage{amsmath}
\usepackage{amsfonts}
\usepackage{color,xcolor}
\usepackage{mathrsfs,amsmath}
\usepackage{amssymb}
\usepackage{grffile}
\usepackage{aas_macros}
\usepackage{array}
\usepackage{url}
\usepackage{longtable}

\usepackage{cases}
\usepackage{longtable}
\usepackage{ulem,array}
\usepackage{multirow}
\usepackage[mathscr]{euscript}

\usepackage{hyperref}
\hypersetup{
  bookmarks=false,
  pdfpagelabels=false,
  hyperfootnotes=false,
  hyperindex=false,
  pageanchor=false,
  colorlinks=true,
  linkcolor=blue,
  citecolor=blue,
  }

\usepackage{multirow}
\usepackage{amssymb}

\usepackage[mathscr]{euscript}

\newcommand{\mathbfit}[1]{\textbf{\textit{#1}}}
\renewcommand{\vec}[1]{\mathbfit{#1}}

\newcommand{\alf}{Alfv$\acute{\text{e}}$n} 

\newcommand{\vsh}{v_{\rm sh}}
\newcommand{\cs}{c_{\rm s}}
\newcommand{\vA}{v_{\rm A}}
\newcommand{\vu}{v_{\rm u}}

\newcommand{\Ma}{\mathcal{M}_A}
\newcommand{\Ms}{\mathcal{M}_{\rm s}}

\DeclareSymbolFont{matha}{OML}{txmi}{m}{it}
\DeclareMathSymbol{\varv}{\mathord}{matha}{29}

\usepackage{array}
\newcolumntype{M}[1]{>{\centering\arraybackslash}m{#1}}
\newcolumntype{N}{@{}m{0pt}@{}}

\begin{document}

\author{Mohamad Shalaby}
\affiliation{Leibniz-Institut f{\"u}r Astrophysik Potsdam (AIP), An der Sternwarte 16, 14482 Potsdam, Germany;
\href{mailto:mshalaby@live.ca}{mshalaby@live.ca}
}

\title{Energy Dissipation in Strong Collisionless Shocks:\\The Crucial Role of Ion-to-Electron Scale Separation in Particle-in-Cell Simulations}

\begin{abstract}
Energy dissipation in collisionless shocks is a key mechanism in various astrophysical environments. Its non-linear nature complicates analytical understanding and necessitate Particle-in-Cell (PIC) simulations. This study examines the impact of reducing the ion-to-electron mass ratio ($m_r$), to decrease computational cost, on energy partitioning in 1D3V (one spatial and three velocity-space dimensions) PIC simulations of strong, non-relativistic, parallel electron-ion collisionless shocks using the SHARP code. We compare simulations with a reduced mass ratio ($m_r = 100$) to those with a realistic mass ratio ($m_r = 1836$) for shocks with high ($\Ma = 21.3$) and low ($\Ma = 5.3$)  \alf~Mach numbers.
Our findings show that the mass ratio significantly affects particle acceleration and thermal energy dissipation. At high $\Ma$, a reduced mass ratio leads to more efficient electron acceleration and an unrealistically high ion flux at higher momentum. At low $\Ma$, it causes complete suppression of electron acceleration, whereas the realistic mass ratio enables efficient electron acceleration. The reduced mass ratio also results in excessive electron heating and lower heating in downstream ions at both Mach numbers, with slightly more magnetic field amplification at low $\Ma$. Consequently, the electron-to-ion temperature ratio is high at low $\Ma$ due to reduced ion heating and remains high at high $\Ma$ due to increased electron heating. In contrast, simulations with the realistic $m_r$ show that the ion-to-electron temperature ratio is independent of the upstream magnetic field, a result not observed in reduced $m_r$ simulations.
\end{abstract}


\shorttitle{Impacts of reduced mass-ratio in PIC simulations of shocks}
\shortauthors{M. Shalaby}

\received{3 August 2024}
\accepted{3 Dec 2024}
\submitjournal{Astrophysical Journal Letters}


\section{Introduction}

Understanding the mechanisms of energy dissipation in strong collisionless shocks is critical for advancing our knowledge of various astrophysical phenomena. These shocks are ubiquitous in the universe, occurring in diverse environments such as supernova remnants, the solar wind, the interstellar medium, and galaxy clusters \citep{Willingale+1996,Helder2009,Brunetti2014,2016Girichidis,2019vanWeeren}. They play a significant role in accelerating cosmic rays, generating magnetic fields, and heating plasmas, which in turn influence the dynamics and evolution of galaxies and galaxy clusters \citep{1987Blandford,guo2008,Gaisser2016,2016Simpson}.

Astrophysical shocks convert bulk kinetic energy into thermal energy, non-thermal particle acceleration, and magnetic field amplification. For example, supernova remnants, driven by the explosive death of massive stars, propagate shock waves into the surrounding interstellar medium, accelerating particles to relativistic speeds and contributing to the galactic cosmic ray population and their dynamical impacts \citep{Bykov+1999,2018Girichidis}. Similarly, shocks in the solar wind, such as those formed at the Earth's bow shock, are crucial for understanding space weather phenomena and their impact on planetary magnetospheres \citep{Simpson+1974,Parker+Zank+2012,Kronberg+2021}. The study of collisionless shocks thus offers insights into fundamental processes governing high-energy astrophysical environments.

The strong non-linearity in collisionless shocks arises from the complex interplay between particles and fields. As particles cross the shock front, they interact with the electromagnetic fields generated by the shock, leading to a feedback loop that drives instabilities and turbulence \citep{Malkov+2001}. These non-linear interactions result in the generation of large-amplitude waves and the formation of fine-scale structures that are crucial for particle acceleration and energy dissipation~\citep{Axford1977}.

The inherently non-linear nature of collisionless shocks poses significant challenges for analytical approaches, necessitating high-resolution simulations that can accurately resolve the dynamics of both ions and electrons \citep{Malkov+2001,Marcowith2016}. Particle-In-Cell (PIC) simulations have become indispensable tools for studying these shocks. PIC simulations provide a first-principles approach by solving the equations of motion for individual particles and self-consistently evolving the electromagnetic fields \citep{buneman+1959,dawson+1962}. This method captures the microphysical processes essential for understanding energy dissipation and particle acceleration in collisionless shocks \citep[e.g.,][]{tristan-mp+2005}.

One of the critical challenges in simulating collisionless shocks is the vast separation of scales between ions and electrons. The ion-to-electron mass ratio, approximately 1836 in nature, results in significant differences in their dynamics and interactions with the shock. To reduce computational costs, a reduced ion-to-electron mass ratio is often used in PIC simulations, lowering the separation between the necessary plasma frequency time scale and the longer ion-plasma frequency time scale. While this approach is justified by computational constraints and has been adopted to study the complex and non-linear physics of these shocks \citep[see, e.g.,][]{Riquelme+2011,Park+2015,Marcowith2016,Xu2020,KumarReville2021,Bohdan+2022,Vanthieghem+2022,Gupta2024}, it raises questions about the accuracy of such simulations in replicating scenarios with realistic mass ratios. Reduced mass ratio simulations can inaccurately represent key processes, such as electron acceleration and heating efficiencies \citep{Bret+2010PhPl,Shalaby+2022ApJ}.

Numerical calculations reveal how upstream kinetic energy is dissipated into thermal and non-thermal energies among various plasma species, along with the mechanisms responsible for this energy partitioning. These factors are interdependent; if simulations with a reduced mass ratio alter physical processes, the results may be nonphysical. In this letter, we focus on the role of the ion-to-electron mass ratio in energy dissipation processes of strong collisionless shocks. By comparing simulations with realistic and reduced mass ratios, we aim to elucidate the importance of accurately modeling ion-to-electron scale separation to capture the true dynamics of collisionless shocks. We investigate the impact of this choice on both the acceleration and heating of ions and electrons in strong parallel non-relativistic shocks in a 1D3V setup.
We utilize the Particle-in-Cell code SHARP-1D3V~\citep{sharp,sharp2,Lemmerz+2023}, which employs higher-order interpolation to couple the Lagrangian and Eulerian parts of the algorithm, reducing numerical heating and maintaining exact momentum conservation. This capability is essential for simulations with realistic mass ratios, ensuring high numerical fidelity over extensive computational runs and avoiding qualitatively incorrect physical results.
The code was initially developed to investigate TeV blazar-driven beam-plasma instabilities in the intragalactic medium~\citep{blazari,blazarii,blazariii,chang:2014,lamberts:2015,linear-paper,bowtiesi,bowtiesii,bowtiesiii,shalabythesis2017,resolution-paper,broderick+2018,sim_inho_18,Vafin+2018,th_inho_20,Lamberts+2022}.

We focus on the case of non-relativistic shocks with high sonic Mach number $\Ms = 365$  and study the impacts on simulations with both low and high \alf ic Mach numbers $\Ma = 5.3$ and $\Ma = 21.3$. At low $\Ma = 5.3$, \citep{Shalaby+2022ApJ} showed a transition in the nature of the instabilities expected at the shock transition and downstream regions, significantly affecting electron-acceleration efficiency. Simulations with realistic mass ratios destabilize wave-modes via the intermediate-scale instability~\citep{sharp2,Shalaby2023,Lemmerz+2024a}, greatly increasing electron-acceleration efficiency compared to cases with suppressed instabilities due to reduced mass ratios. At high $\Ma = 21.3$, where the instability is stabilized, it is questioned whether reduced mass ratio simulations can faithfully mimic the physics observed in realistic mass ratio simulations. In this letter, we demonstrate that for high $\Ma = 21.3$, a reduced mass ratio of $m_r = 100$ leads to nonphysical results in both acceleration and heating, especially for electron species in the downstream region of the shock. Our findings indicate that the impacts are more severe than previously anticipated; reduced mass ratios result in incorrect acceleration efficiencies and plasma heating for both electrons and ions at both low and high values of $\Ma$. This work underscores the necessity of using realistic mass ratios in PIC simulations to improve our understanding of astrophysical shocks.

The letter is organized as follows. In Section~\ref{sec:setup}, we describe the numerical setup for our simulations. 
We discuss the expected scales for unstable wave modes driven during shock formation and how these are related to the Larmor radius of particles in Section~\ref{sec:scales}. 
Section~\ref{sec:results} presents the impacts of the reduced mass ratio on magnetic field amplification, thermal and non-thermal energy dissipation at high and low $\Ma$. We conclude and summarize our findings in Section~\ref{sec:conclusion}. Throughout this letter, we assume the SI system of units.

\begin{figure*}[!ht]
\includegraphics[width=19cm]{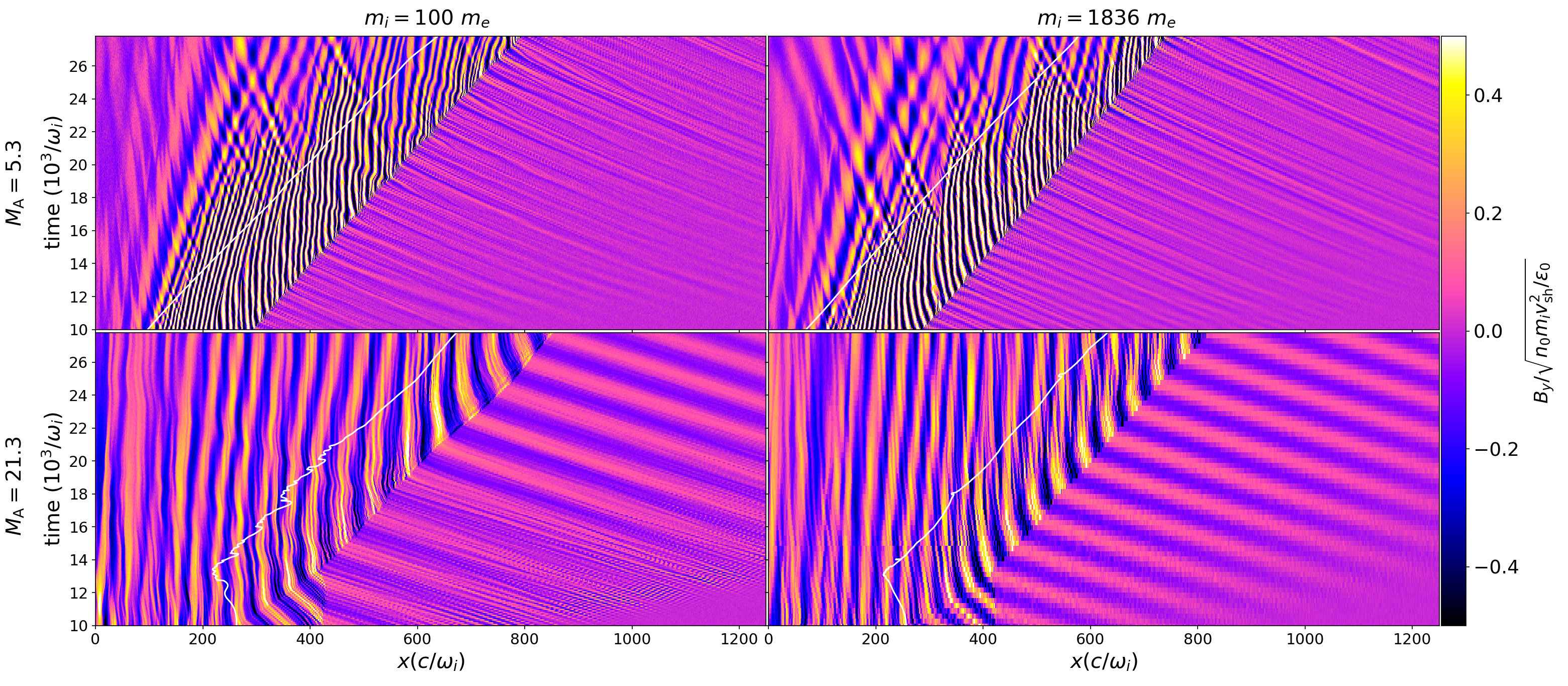}
\caption{\label{fig:By}%
Amplification and spatial structure of the perpendicular magnetic field component $B_y$ in various simulations. The top panels show simulations with $\Ma = 5.1$, and the bottom panels show simulations with $\Ma = 21$. The left panels present simulations with a reduced ion-to-electron mass ratio $m_r = 100$, while the right panels show simulations with the realistic mass ratio $m_r = 1836$.
The normalization is such that the square of this value represents the fraction of upstream shock kinetic energy density converted into the magnetic energy density of the $B_y$ magnetic field component. In all cases, large-scale \alf~waves are clearly generated near the shock front, with wavelengths approximately given by $\lambda \sim 2 \pi \Ma c / \omega_i$, which is consistent with the scale expected from a streaming instability gyroscale wave mode driven by the incoming upstream cold plasma penetrating through the downstream denser electron-ion plasma (see Section~\ref{sec:scales}). The region to the left of the white lines indicates what we refer to as the downstream region of the shock, which is approximately $200 c / \omega_i$ away from the shock front, evidenced by a clear transition in the field amplitude in all panels.
This shows that in all simulations, roughly 10\% of the upstream plasma kinetic energy is converted into magnetic field energy, seemingly independent of the far upstream magnetic energy density.
}
\end{figure*}

\section{Simulation setup}
\label{sec:setup}

Simulations are conducted using the Particle-in-Cell code SHARP-1D3V \citep{sharp,sharp2}, which utilizes higher-order interpolation to integrate the Lagrangian and Eulerian components of the algorithm. This approach significantly reduces numerical heating over extended time steps while preserving exact momentum conservation \citep{sharp}. Such precision is vital for simulations with realistic mass ratios, which require numerous computational steps. High numerical fidelity throughout extensive simulations is crucial \citep{resolution-paper,sim_inho_18}, as most standard PIC algorithms \citep{birdsall+1980,hockney-eastwood,lipatov+2002} in existing codes struggle to accurately simulate realistic mass ratios and long integration periods without introducing excessive numerical heating, potentially resulting in incorrect physical outcomes \citep{sharp}.

The simulations are conducted in the contact discontinuity rest frame, as described in~\citep{Shalaby+2022ApJ}. In this setup, a shock is formed by reflecting an electron-ion plasma moving leftward (upstream plasma) at \( x=0 \), creating a shock from the interaction of two identical plasmas moving in opposite directions.
The upstream average velocity for both electrons and ions is \( \vu = -0.1c \), where \( c \) is the speed of light. We include a fixed parallel large-scale magnetic field, \( \vec{B}_0 = B_0 \hat{x} \).

In all simulations, we resolve the upstream electron skin depth with 10 cells, and each computational cell contains 200 particles per species in the far-upstream region. The time step is \( 0.045 \omega_p^{-1} \), where \( \omega_p = \sqrt{\omega_e^2 + \omega_i^2} \) is the total plasma frequency. The electron and ion plasma frequencies are \( \omega_e = \sqrt{e^2 n_e / \epsilon_0 m_e} = \omega_p \sqrt{m_r / (m_r+1)} \) and \( \omega_i = \sqrt{e^2 n_e / \epsilon_0 m_i} = \omega_p / \sqrt{m_r+1} = \sqrt{m_r} \omega_e \), respectively, with \( e \) being the elementary charge, \( m_r = m_i / m_e \) the ion-to-electron mass ratio, and \( \epsilon_0 \) the vacuum permittivity.

We initialize equal number densities in the upstream region for both electrons and ions, i.e., \( n_e = n_i = n_0 \). The shock speed in the rest frame of the upstream plasma, \( \vsh \), depends on the shock compression ratio, \( R = n / n_0 \) with \( n \) being the downstream average number density, as follows:
\begin{eqnarray}
\vsh = \vu \left[ 1 + \frac{1}{R-1} \right] = \frac{R}{R-1} \vu .
\label{eq:vshvu}
\end{eqnarray}

We follow the definitions given in \citep{Shalaby+2022ApJ}.
The {\alf}ic Mach number is \( \Ma = \vsh / \vA \), where the ion \alf~speed is \( \vA = B_0 / \sqrt{n_0 m_i / \epsilon_0} \). The sonic Mach number is \( M_s = \vsh / \cs \), with the sonic speed \( \cs = \sqrt{\Gamma_{\rm ad} k_B (T_e + T_i) / m_i} \), where \( \Gamma_{\rm ad} = 5/3 \) is the adiabatic index, \( k_B \) the Boltzmann constant, and \( T_e \) (\( T_i \)) the electron (ion) upstream isotropic temperature. We fix the upstream electron and ion temperatures such that \( k_B T_e = k_B T_i = 4 \times 10^{-8} m_i c^2 \). Thus, \( \Ms \approx 274 \times R / (R-1) \) in all simulations.

\section{Collisionless shocks: Theoretical considerations}

\label{sec:scales}

Particles with larger momenta can resonantly interact with longer wavelength magnetic perturbations, and thus we compute, below, relevant length scales and relate them to particle momenta.
As done in simulations, we measure these in terms of the upstream ion plasma skin-depth, $d_i$ and ion-\alf~speed defined in terms of upstream number density $n_0$ and magnetic field strength $B_0$.

\subsection{Normalized gyro-radii}

The gyro-radius of a particle of species $s=i(e)$, i.e., ions (electrons), is given by
\begin{eqnarray}
r_s &=&  \frac{  v_{s,\perp}}{\Omega_s} =
\frac{\gamma_s m_s  v_{s,\perp}}{ m_i \Omega_i}
=
\frac{p_{s,\perp}}{ m_i v_A} d_i
\approx
\frac{1}{2}\frac{p_{s}}{ m_i \vA} d_i
\\
&& \Rightarrow 
\frac{r_s}{d_i}  =
\frac{1}{2}\frac{p_{s}}{ m_i \vsh} \Ma.
\end{eqnarray}
where $\Omega_s$ is the relativistic gyrofrequency, and $\Ma$ is the upstream \alf~Mach number.
We used $\Omega_i = \omega_i \vA/c = \vA/d_i $ where $d_i= c/\omega_i$ is the upstream ion skin-depth, and assumed isotropic plasmas, i.e., $p_{s,\perp} = p_{s,\parallel} \approx p_s/2$. It is important to note that we computed the gyro-radius in terms of the upstream magnetic field, i.e., we did not incorporate magnetic field amplification in the downstream. In our 1D3V simulations, the geometry implies that the large-scale parallel magnetic field remains constant.

\subsection{Wavelength for unstable {\alf}-Waves}
The large-scale waves in the downstream region are {\alf}-waves and are driven due to streaming instability~\citep{kulsrud+1969,Shalaby2023}.
We first estimate this near the shock front towards the downstream region, the wave modes are such that
\begin{eqnarray}
\frac{ k_G d_i }{ \sqrt{R}}&=& 
\frac{2 \pi d_i}{ \sqrt{R} \lambda^{d}_G}
= \frac{1}{\sqrt{R} \vsh/\vA -1}
\\
&&
\Rightarrow~
\frac{ \lambda^{d}_G}{d_i}
= 
2 \pi \left[\Ma - \frac{1}{\sqrt{R}}  \right]
\approx
2 \pi \Ma
\label{eq:lg_d}
\end{eqnarray}
where $R$ is the average compression ratio in the downstream region, and $\lambda^{d}_G$ is the wavelength of the gyroscale wavemodes in the downstream region.
Similarly, ions escape from the downstream region towards upstream region, i.e., with speed $v \geq \vsh$,  excite \alf~wave near the shock front towards the upstream region\footnote{While the growth rate depends on the density of these ions, the wavelength of the fastest unstable gyro-scale does not in the limit of low density CR ion escaping, which is the case here \citep{Shalaby2023}.} such that 
\begin{eqnarray}
k_G d_i &=& 
\frac{2 \pi d_i}{ \lambda^{u}_G}
< \frac{1}{ \vsh/\vA -1}
\\
&&
\Rightarrow~
\frac{ \lambda^{u}_G}{d_i}
> 2 \pi \left[\Ma - 1 \right]
\approx
2 \pi \Ma.
\label{eq:lg_u}
\end{eqnarray}
Where $\lambda^{u}_G$ is the wavelength of the gyroscale wavemodes in the upstream region.
Equations\eqref{eq:lg_d} and \eqref{eq:lg_u} show that the expected most unstable wavelength is the same at both upstream and down stream regions. This is clearly manifested in our simulations as can be seen in Figure~\ref{fig:By}.
It is important to also note here that when the flux of the escaping ions towards the upstream region is high, this can destabilize Bell unstable wave-modes \citep{Bell2004}. However, above we focus on the initial low flux modes where the streaming instability is the dominant instability at scales larger than the ion-skin depth~\citep{Zweibel2017}.

\subsection{Wavelength for intermediate-scale unstable waves}

When the intermediate-scale instability operates in the downstream region of the shock, i.e., if $\Ma < \sqrt{m_i/m_e}/ 2\sqrt{R}$,  it destabilizes wave-modes where the fastest growth~\citep{sharp2,Shalaby2023} occurs for wave-mode, $k_{\rm I}$, i.e., 
\begin{eqnarray}
\frac{ \lambda_{\rm I}}{d_i}
= \frac{ 2 \pi }{ k_{\rm I} d_i }
=  \frac{ 2 \pi }{ \bar{k}_{\rm I}}.
\end{eqnarray}

In \citet{sharp2}, it was estimated that $\bar{k}_{\rm I} \approx R \Ma$\footnote{The peak growth occurs at 
$k_{\rm I} d_i^d \approx \vsh/\vA^d~ \Rightarrow~
k_{\rm I} d_i /\sqrt{R}   \approx  \sqrt{R} \Ma \Rightarrow~ \bar{k}_{\rm I} \approx R \Ma
$. Here, the superscript $d$ refers to quantities measured in the downstream region of the shock.}, however, as noted there, this is a rough approximation. More accurate values can be obtained by solving the linear dispersion relation or a simpler polynomial equation given in Equation (2.7) of \citet{Shalaby2023}.
Thus, electrons with $r_s \geq \lambda_{\rm I}$ can efficiently interact with the unstable modes, leading to
\begin{eqnarray}
r_s \geq \lambda_{\rm I}
~\Rightarrow~
\frac{p_{s}}{ m_i \vsh} 
\geq \frac{ 4 \pi }{ \Ma \bar{k}_{\rm I} } \approx
\frac{ 4 \pi }{ R \Ma^2 } 
\label{eq:rg_I}
\end{eqnarray}

It is shown in the simulation of \citet{sharp2} that the unstable modes very quickly  inverse cascade to larger scales in the 1D3V geometry, and thus we expect electrons with $r_s > \lambda_{\rm I} $to efficiently interact with these intermediate-scale unstable wave-modes.
Power in the perpendicular magnetic perturbation at scales $\geq \lambda_{\rm I} $ is very pronounced in the simulation where intermediate-scale modes are unstable in line with what is seen in the bottom panel of Figure 4 in \cite{Shalaby+2022ApJ}.
 
The intermediate-scale modes are the shortest relevant waves expected to grow and interact with both ion and electron species. The wavelength is $\sim 2 \pi M_A d_i /\sqrt{R} = 2 \pi \sqrt{m_r+1} M_A d_e /\sqrt{R} $, and thus it is resolved by $20 \pi \sqrt{m_r+1} M_A /\sqrt{R}$ cells in our simulations. This electromagnetic wave mode is also carried by approximately $4000 \pi \sqrt{m_r+1} M_A /\sqrt{R}$ ions and $4000 \pi \sqrt{m_r+1} M_A /\sqrt{R}$ electrons.
Therefore, the observed severe impacts of using an artificially low mass ratio are not due to limitations in the spatial or velocity resolution of the simulations.

\begin{figure*}[!ht]
\includegraphics[width=19cm]{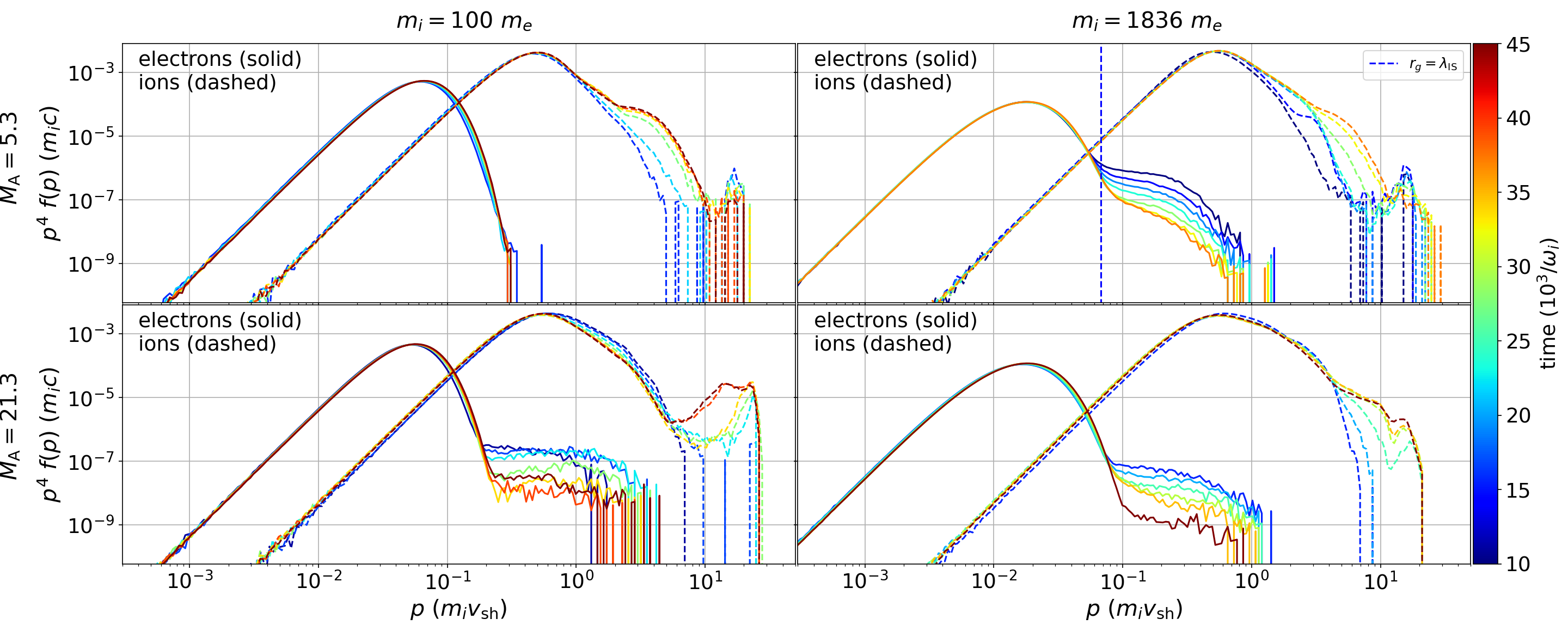}
\caption{\label{fig:spectra}%
The time evolution of particle spectra (solid lines: electrons; dashed lines: ions) at the downstream region defined as the region to the left of white curved in Figure~\ref{fig:By}. All spectra are computed from particle momenta in the contact-discontinuity (simulation) frame. The momentum is normalized to the upstream kinetic energy in the shock rest-frame, with $\vsh$ computed from Equation~\eqref{eq:vshvu} assuming a compression ratio of $R = 4$.
The top right panels show results from the simulation that can destabilize intermediate-scale unstable wave modes. The vertical blue line indicates the particle momenta for which the particle gyro-radius is in resonance with the fastest growing modes (Equation~\eqref{eq:rg_I}).
At a low \alf ic Mach number $\Ma = 5.3$, the reduced mass ratio leads to complete suppression of electron acceleration, while very efficient electron acceleration is observed with the realistic mass ratio, enabling destabilization of the intermediate-scale instability. Ion acceleration, however, is similar in both simulations.
At a high \alf ic Mach number $\Ma = 21.3$, the reduced mass ratio results in much more efficient electron acceleration compared to the simulation with the realistic mass ratio, indicating that the mechanism for such acceleration is also sensitive to the mass ratio. This sensitivity is also observed in ion acceleration, where the reduced mass ratio leads to an nonphysically high flux of ions at high momentum compared to the simulation with the realistic mass ratio.
}
\end{figure*}

\section{Impacts of reduced mass-ratio in simulations}
\label{sec:results}

In this section, to determine the impacts of reduced electron-ion scale separation, i.e., the impacts of reduced \(m_r\), we compare the results of simulation with two \alf ic Mach numbers, \(\Ma = 5.3\) and \(\Ma = 21.2\), each with two different ion-to-electron mass ratios: \(m_r = 100\) and \(m_r = 1836\).
We vary the value of $\Ma$, by varying the upstream magnetic field $B_0$ while keeping the upstream plasma speed $\vu=0.1c$ constant.
All simulations have the same sonic Mach number\footnote{
Fixing the far upstream $\Ms$ (with $T_e=T_i$, and fixed $\vu$) imply that the thermal spread in ion velocity remains the same in all simulations, while the thermal spread in the electron velocity is larger in simulation with reduced $m_i$ compared to simulations with the realistic $m_i$.}  $\Ms \approx 274 \times R/(R-1)$. 
If we assume a compression ratio of $R=4$, the sonic Mach number in all simulations is $\Ms = 365$ and the {\alf}ic Mach numbers are $\Ma = 5.3$ and $21.3$.

Below, we discuss the differences and similarities across various simulations, and we make visualization movies of the full evolution of different quantities for all simulations publicly available at
\href{https://mohamadshalaby.github.io/Pshock_mr}{mohamadshalaby.github.io$\slash$Pshock\_mr}.

\subsection{Magnetic field amplification}

In the magnetohydrodynamics (MHD) limit for parallel shocks, the upstream kinetic energy is fully dissipated into thermal energy in the downstream region of the shock, without any magnetic field amplification expected~\citep[see, e.g.,][]{boyd}. However, in the collisionless limit, the perpendicular magnetic field is amplified due to driven instabilities. These unstable electromagnetic modes are responsible for shock formation in the collisionless case, meaning that a fraction of the upstream kinetic energy is necessarily converted into magnetic energy in our simulations.

In this section, we examine the amplification and spatial structure of the perpendicular magnetic field, using the component $B_y$ across different simulations, as shown in Figure~\ref{fig:By}. The top panels present simulations with an \alf ic Mach number $\Ma = 5.1$, while the bottom panels depict simulations with $\Ma = 21$. The left panels feature simulations utilizing a reduced ion-to-electron mass ratio $m_r = 100$, whereas the right panels showcase simulations with the realistic mass ratio $m_r = 1836$. 

In all simulations, large-scale \alf~waves are evidently generated near the shock front, with wavelengths approximately given by $\lambda \sim 2 \pi \Ma c / \omega_i$. This observation aligns with the expected scale from a streaming instability gyroscale wave mode driven by the incoming upstream cold plasma penetrating the downstream denser electron-ion plasma (see Section~\ref{sec:scales}). The region to the left of the white lines is identified as the downstream region of the shock, located approximately $200 c / \omega_i$ away from the shock front, which is marked by a noticeable transition in the field amplitude across all panels.

Normalization in the figure is such that the square of the displayed values represents the fraction of upstream shock kinetic energy density converted into the magnetic energy density of the $B_y$ magnetic field component. This illustrates that in all simulations, approximately 10\% of the upstream plasma kinetic energy is converted into magnetic field energy. This converted fraction appears to be independent of the far upstream magnetic energy density.

\subsection{Evolution of Downstream Particle Spectra}

\begin{figure}[!ht]
\includegraphics[width=8.8cm]{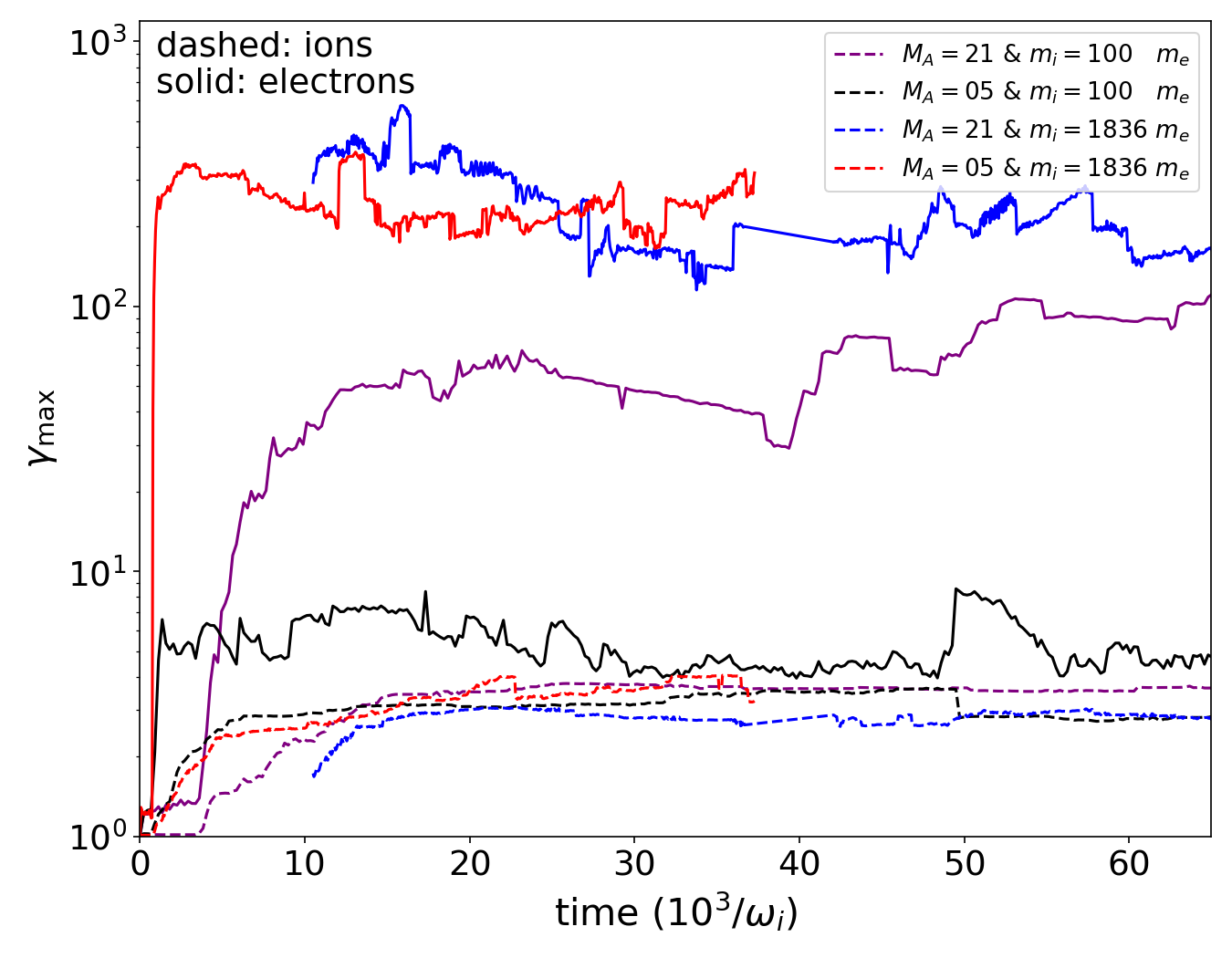}
\caption{\label{fig:GamMax}%
The time evolution of the maximum energy for ions and electrons normalized by their rest-mass energy, i.e.,  $\gamma_{\rm max}$, at the shock front region which is defined as 400 $c/\omega_i$ region centered at the wave intensity jump locations shown in Figure~\ref{fig:By}.
This shows that the maximum electron energy increases rapidly during shock formation in all simulations, with some simulations showing a slower increase or saturation afterward. For ions, the initial energy increase is slower than for electrons and appears consistent across simulations with the same Alfvénic Mach number, followed by slower growth with less oscillatory behavior.
}
\end{figure}

In this section, we examine the time evolution of particle spectra. These results are presented in Figure~\ref{fig:spectra}, where solid lines represent electrons and dashed lines represent ions. The spectra are computed from particle momenta in the contact-discontinuity (simulation) frame in the downstream region, i.e., the region to the left of the white curve in Figure~\ref{fig:By}. For consistency, the momentum is normalized to the upstream kinetic energy in the shock rest frame. The shock velocity \(\vsh\) is calculated (for \(R = 4\)) using Equation~\eqref{eq:vshvu}.

The top right panel of the figure presents results from simulations that destabilize intermediate-scale unstable wave modes. A vertical blue line in these panels marks the particle momenta at which the particle gyro-radius resonates with the fastest growing modes, as described by Equation~\eqref{eq:rg_I}.

At a low \alf ic Mach number of \(\Ma = 5.3\), the results demonstrate a significant effect of the mass ratio on electron acceleration. Specifically, simulations with a reduced ion-to-electron mass ratio exhibit complete suppression of electron acceleration. In contrast, simulations with the realistic mass ratio show very efficient electron acceleration. This efficiency is attributed to the ability of the realistic mass ratio to destabilize the intermediate-scale instability~\citep{Shalaby+2022ApJ}. Interestingly, ion acceleration remains quite similar across both mass ratio simulations, indicating that the mass ratio primarily affects electron dynamics at this lower Mach number.

The figure further explores the impact at a higher \alf ic Mach number of \(\Ma = 21.3\). The reduced mass ratio leads to much more efficient electron acceleration compared to simulations using the realistic mass ratio. This finding suggests that the mechanism driving electron acceleration is highly sensitive to the mass ratio, even though the intermediate-scale modes are not destabilized in either simulation. This mass ratio sensitivity is also evident in ion acceleration. The reduced mass ratio results in a nonphysically high flux of ions at high momentum, a phenomenon not observed in simulations with the realistic mass ratio. This discrepancy underscores the importance of using realistic mass ratios to obtain accurate representations of particle acceleration processes in both ions and electrons in the high \(\Ma\) cases.

From Figure~\ref{fig:spectra}, it is difficult to infer the behavior of the maximum particle energy. Since most of the acceleration likely occurs in the shock front region, we plot the evolution of the highest energy for both electrons and ions near the shock front in Figure~\ref{fig:GamMax}. We define the shock front region as a 400 $c/\omega_i$ area centered at the wave intensity jump locations shown in Figure~\ref{fig:By}.

This analysis shows that for electrons, the maximum energy increases very rapidly during shock formation. In some simulations, such as $\Ma=21$ \& $m_r=100$ and $\Ma=5$ \& $m_r=1836$, a slower energy increase is observed after this initial phase, while in others, the increase is either even slower or reaches saturation. However, it is difficult to determine which scenario applies to each simulation without a much longer time evolution. We leave this important point for future investigations.

For ions, on the other hand, the rapid energy increase during shock formation is slightly slower compared initial rate for electrons. This initial rate is very similar in simulations with the same Alfvénic Mach number, regardless of the mass ratio, due to the slower growth rates of waves that scatter ions compared to those affecting electrons. Since this initial rapid rate appears independent of the mass ratio, the slower increase compared to that of electrons is not due to the higher mass of ions. Following this rapid phase, the maximum ion energy shows a much slower growth, similar to that of electrons, but with far less oscillatory behavior.
In various simulations, the initial rates of energy increase for electrons and ions differ, suggesting that distinct acceleration mechanisms are at play initially. Investigating the precise details and modeling these mechanisms is left for future work.

We use the particle spectra to infer the fraction of upstream kinetic energy dissipated into thermal and non-thermal energy for both ions and electrons in the simulations in the next sections.

\subsection{Dissipation into thermal energy}

\begin{figure*}[!ht]
\includegraphics[width=19cm]{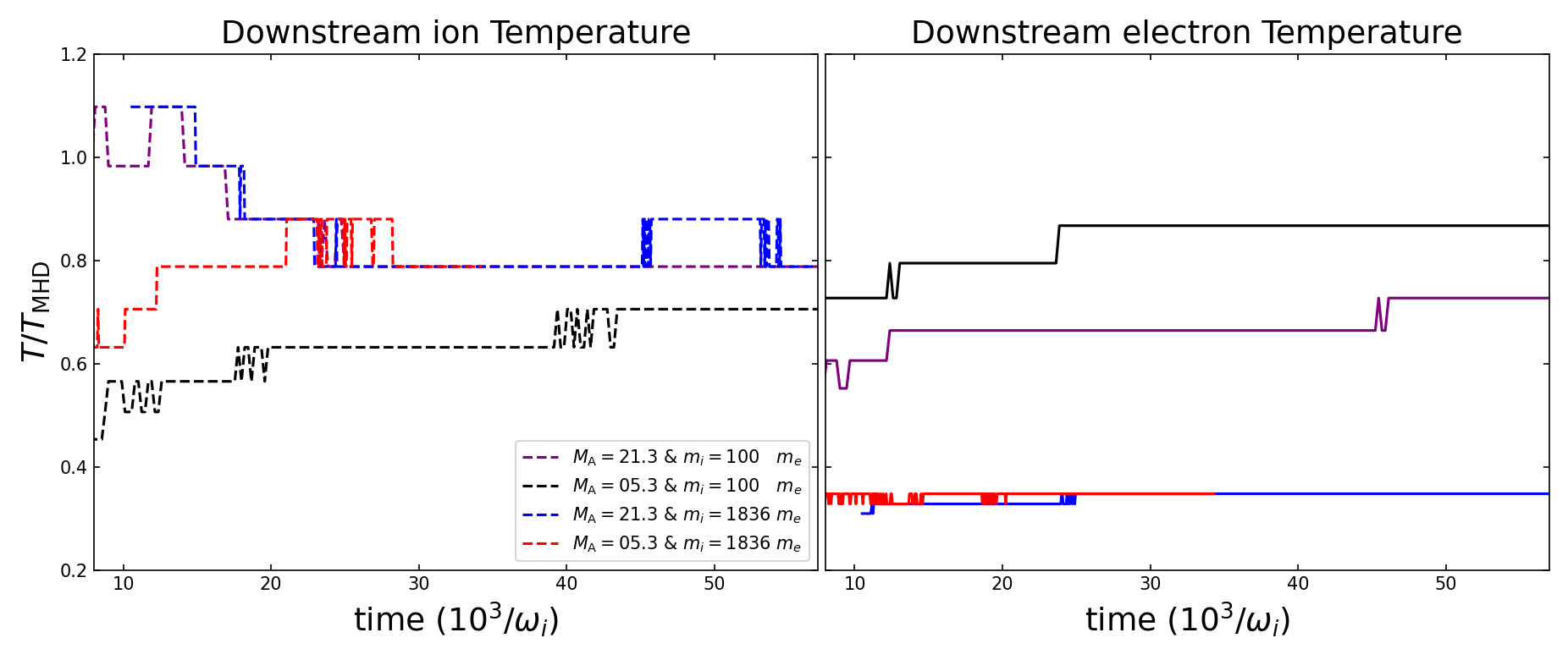}
\caption{\label{fig:Dn_temps}%
The time evolution of downstream ion (left panel) and electron (right panel) temperatures in various simulations. The values are normalized with $T_{\rm MHD}$, which is computed using Equation~\eqref{eq:Tmhd} assuming a compression ratio of $R=4$. Initially, when the compression is less than $4$, the effective value of $T_{\rm MHD}$ is larger than that used in the normalization. Consequently, the downstream temperatures of the ions might initially exceed $T_{\rm MHD}$ used in the normalization. This is observed in the $\Ma=21$ simulations for $t \omega_i < 15 \times 10^3$ (see bottom panels of Figure~\ref{fig:By}).
This figure shows that using a reduced mass ratio leads to excessive heating for electrons at both low and high values of $\Ma$. At $\Ma=5.3$, this also results in lower heating in the downstream ions compared to the simulation with the same physical parameters and a realistic mass ratio. In all simulations, almost all the thermal energy is carried by ions (even if $T_e=T_i$).
Simulations with a realistic mass ratio convert about 78\% of the upstream kinetic energy into thermal energy in the downstream region, primarily carried by ions. In contrast, simulations with a reduced $m_r$ recover this percentage only at high \alf ic Mach numbers, and it is slightly lower (72\%) at low \alf ic Mach numbers, which dissipate more energy in magnetic fields as seen in Figure~\ref{fig:By}.
}
\end{figure*}

One can find an upper limit to how much energy can end up into heating using the MHD jump condition, that is if all the kinetic energy of the upstream plasma is dissipated into thermal energy in the downstream rest frame, the resulting expected temperature for electrons or ions is given by \citep{boyd}
\begin{eqnarray}
k_B T_{MHD} = \frac{3}{32} m_i \vsh^2
=  \frac{3R^2}{32(R-1)^2} m_i \vu^2
\label{eq:Tmhd}
\end{eqnarray}
where, we assumed fully ionized electron-ion plasma, and $m_i$ is the ion mass, and $\vsh$ is the shock speed in the rest-frame of the shock. $\vsh$ is related to upstream plasma speed in the contact discontinuity rest frame, $\vu$, and depends on the shock compression ration, $R$, as given in Equation~\eqref{eq:vshvu}.
We note that the MHD jump conditions give a downstream temperature of \(2 T_{\text{MHD}}\). Since the MHD approximation assumes equal electron and ion temperatures/pressures, and the MHD pressure is the sum of both \citep[Equation 6.14]{Braginskii_1965}, we define \(T_{\text{MHD}}\) (as given in Equation~\eqref{eq:Tmhd}) to directly compare with the electron and ion temperatures found in simulations.

To measure a meaningful temperature for particles with both thermal and non-thermal components in their momentum distribution, one cannot directly compute this by calculating the standard deviation around the average speed. However, one can determine the temperature for a Maxwell-J{\"u}ttner distribution that fits the low-energy part, which can be directly inferred from the momentum where the particle spectra (shown in Figure~\ref{fig:spectra}) is maximized. This is shown mathematically in Appendix A of \citep{Shalaby+2022ApJ}.

\begin{figure}
\includegraphics[width=8.8cm]{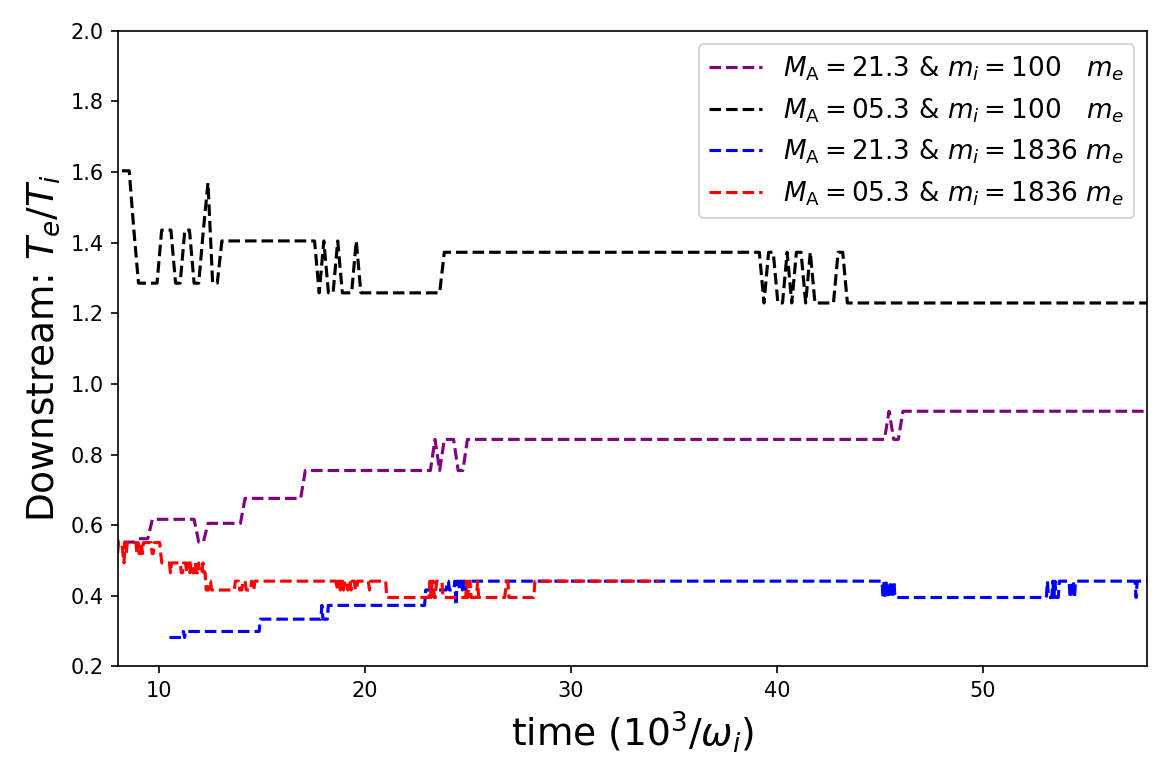}
\caption{\label{fig:TeTi}%
The time evolution of the ratio of downstream electron-to-ion temperatures in various simulations. In the case of a low {\alf}ic Mach number ($\Ma = 5.3$) with a reduced mass ratio (represented by the black curve), the high electron-to-ion temperature ratio is due to significantly less heating of ions, as shown in the left panel of Figure~\ref{fig:Dn_temps}. At a high $\Ma$ ($\Ma = 21.3$), the ratio remains higher at the reduced mass ratio because of the significantly more heating of electrons compared to the simulation with the same physical parameters but a realistic mass ratio. 
This figure highlights an important result for this work: the ion-to-electron temperature ratio is independent of the value of $\Ma$ (or more precisely, independent of the upstream magnetic field value). This result cannot be deduced from simulations with a reduced mass ratio.}
\end{figure}

Figures~\ref{fig:Dn_temps} and~\ref{fig:TeTi} provide insights into the thermal dynamics of electrons and ions in shock simulations with varying Mach numbers and mass ratios.
Figure~\ref{fig:Dn_temps} presents the time evolution of downstream ion (left panel) and electron (right panel) temperatures across various simulations. The temperatures are normalized using $T_{\rm MHD}$, computed with Equation~\eqref{eq:Tmhd} assuming a compression ratio of $R = 4$. Initially, due to the compression being less than 4, the effective $T_{\rm MHD}$ exceeds the normalization value, causing downstream ion temperatures to initially surpass $T_{\rm MHD}$, especially in the $\Ma = 21$ simulations for $t \omega_i < 15 \times 10^3$ (see bottom panels of Figure~\ref{fig:By}).

The results indicate that using a reduced mass ratio leads to excessive heating of electrons at both low and high $\Ma$. At $\Ma = 5.3$, this also results in lower heating of downstream ions compared to simulations with the same physical parameters but a realistic mass ratio.
In all simulations, almost all the thermal energy is carried by ions, which is true even when $T_e = T_i$. Simulations with a realistic mass ratio convert approximately 78\% of the upstream kinetic energy into downstream thermal energy, primarily carried by ions. In contrast, simulations with a reduced mass ratio recover this percentage only at high \alf ic Mach numbers, and it is slightly lower (72\%) at low \alf ic Mach numbers, where more energy is dissipated into magnetic fields as can be inferred from Figure~\ref{fig:By} (top-left).

Figure~\ref{fig:TeTi} shows the time evolution of the downstream electron-to-ion temperature ratio in various simulations. In the case of a low \alf ic Mach number ($\Ma = 5.3$) with a reduced mass ratio (represented by the black curve), the high electron-to-ion temperature ratio is due to significantly less heating of ions, as shown in the left panel of Figure~\ref{fig:Dn_temps}. At a high $\Ma$ ($\Ma = 21.3$), the ratio remains higher with the reduced mass ratio because of significantly more heating of electrons compared to simulations with the same physical parameters but a realistic mass ratio.
A critical conclusion from this figure is that the ion-to-electron temperature ratio is independent of the Mach number or, more precisely, the upstream magnetic field value. This independence is a significant observation because it highlights a limitation of simulations using reduced mass ratios, which can not accurately represent the thermal dynamics observed with realistic mass ratios.

In summary, these figures underscore the sensitivity of electron and ion heating to the mass ratio at both high and low values of $\Ma$. This insight is crucial for understanding the thermal behavior of particles in shock environments and evaluating the accuracy of simulation models.

\subsection{Dissipation into non-thermal energy: acceleration efficiency in simulations}

\begin{figure}[!ht]
\includegraphics[width=8.8cm]{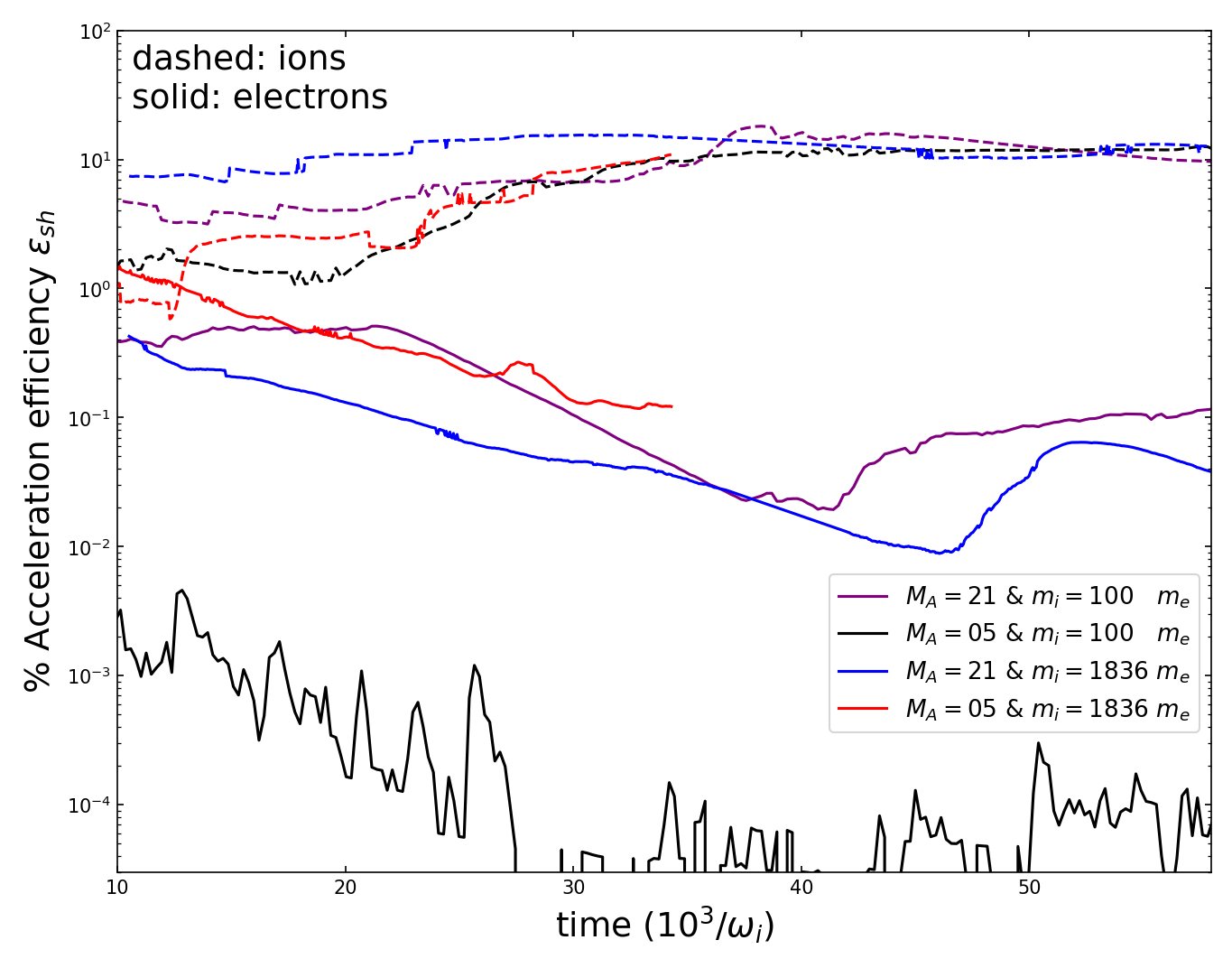}
\caption{\label{fig:efficiency}%
The time evolution of the acceleration efficiency, i.e., the ability of various shocks to dissipate upstream kinetic energy into non-thermal ion (shown as dashed curves) and electron (shown as solid curves) energies. As shown in Figure~\ref{fig:spectra}, in the low $\Ma$ simulation with the reduced mass ratio, electron acceleration is very inefficient. However, at the same $\Ma$ with the realistic mass ratio, the efficiency is the highest among all simulations, including those with higher $\Ma$. At higher $\Ma$, the acceleration efficiency for electrons is systematically greater in the simulation with the reduced mass ratio. While the percentage of non-thermal energy in ions is approximately the same for both reduced and realistic mass ratios, the difference in how this energy is distributed at different momenta is manifested in the lower panels of Figure~\ref{fig:spectra}. In all simulations, about 11-12\% of the upstream kinetic energy is dissipated into non-thermal energy in ions, independent of the value of $\Ma$. This is consistent with the hybrid simulations of \citet{Caprioli2014a}. On the other hand, less than 1\% is dissipated into non-thermal energy in electrons.
This shows that the efficiency especially that of electrons evolves over time, contrary to what is incorrectly inferred from simulations run for much shorter physical times in the literature.
}
\end{figure}

In this section, we examine the time evolution of the acceleration efficiency, which is the ability of various shocks to dissipate upstream kinetic energy into non-thermal ion and electron energies. Similar to the approach taken by \citet{Xu2020} and \citet{Shalaby+2022ApJ}, we consider the energy in the particle distribution with momentum \( p_s > 5 p_{s,\rm max} \), i.e., the energy in particles with momentum 5 times larger than the momentum where the particle distribution is maximized, to be the non-thermal energy dissipated. That is, the non-thermal energy \( E_{s,\rm NT} = E(p_s > 5 p_{s,\rm max}) \), and we define the percentage (\%) acceleration efficiency \( \epsilon_{\rm sh} = 100 \times E_{s,\rm NT}/E_{\rm sh} \), where \( s \) is the species (i.e., electrons or ions), and \( E_{\rm sh} \) is the upstream kinetic energy of the shock.
We show in Figure~\ref{fig:efficiency} the time evolution of the acceleration efficiency, the dashed curves represent the non-thermal ion energies, while the solid curves represent the non-thermal electron energies.

In the low \( \Ma \) simulation with the reduced mass ratio, we observe that electron acceleration is very inefficient. This can be seen from the significantly lower values of the solid black curve. Conversely, when the same low \( \Ma \) simulation is conducted with a realistic mass ratio, the acceleration efficiency for electrons becomes the highest among all simulations, even surpassing those with higher \( \Ma \). This result suggests that the mass ratio plays a crucial role in determining the efficiency of energy transfer to electrons, a result that directly arises from driving the intermediate-scale unstable modes in the simulation with a realistic mass ratio \citep{Shalaby+2022ApJ}.

At higher \( \Ma \) values, the trend reverses. Here, the acceleration efficiency for electrons becomes systematically greater in the simulation with the reduced mass ratio. This suggests that the mass ratio has a different impact on electron acceleration efficiency depending on the shock Mach number. Despite these variations in electron acceleration, the percentage of non-thermal energy in ions remains approximately the same for both reduced and realistic mass ratios. This is evident from the dashed curves, which show little variation between different mass ratio simulations. The lower panels of Figure~\ref{fig:spectra} further illustrate how this energy is distributed across different momenta, highlighting the subtle differences in ion energy distribution.

Across all simulations, we find that about 11-12\% of the upstream kinetic energy is dissipated into non-thermal energy in ions, regardless of the \( \Ma \) value. This finding is consistent with the hybrid simulations conducted by \citet{Caprioli2014a}. On the other hand, less than 1\% of the upstream kinetic energy is dissipated into non-thermal energy in electrons, underscoring the relatively low efficiency of electron acceleration compared to ions in these shock scenarios.

These observations provide valuable insights into the dynamics of shock-driven particle acceleration and the distinct roles played by ions and electrons in this process. The efficiency differences influenced by the mass ratio and Mach number highlight the complex interplay between these parameters in determining the overall acceleration efficiency.

\section{Conclusion}
\label{sec:conclusion}

This letter investigates the impact of the ion-to-electron mass ratio on energy dissipation in collisionless shocks through simulations with varying \alf ic Mach numbers ($\Ma$). We observe distinct differences in shock dynamics, particularly in particle acceleration and thermal energy dissipation.
Our key findings can be summarized as follows:

\begin{enumerate}
    \item \textit{Magnetic Field Amplification}: Both reduced and realistic mass ratio simulations exhibit large-scale \alf~ waves near the shock front, converting approximately 10\% of the upstream plasma kinetic energy into magnetic field energy. This conversion appears independent of the upstream magnetic energy density.

    \item \textit{Thermal Energy Dissipation}: The downstream ion and electron temperatures exhibit distinct dependencies on the mass ratio. Realistic mass ratio simulations consistently show that about 78\% of the upstream kinetic energy is converted into downstream thermal energy, primarily carried by ions. Reduced mass ratio simulations, however, display excessive electron heating and less efficient ion heating at lower Mach numbers, leading to a slightly lower overall thermal energy conversion (72\%). The downstream electron-to-ion temperature ratios reveal crucial insights. At low Mach numbers, a reduced mass ratio results in a high electron-to-ion temperature ratio due to insufficient ion heating. At high Mach numbers, this ratio remains elevated in reduced mass ratio simulations due to disproportionate electron heating. Notably, the electron-to-ion temperature ratio appears independent of the upstream magnetic field, a result not deducible from reduced mass ratio simulations alone.

    \item \textit{Particle Acceleration -- Non-Thermal Energy Dissipation}: The efficiency of electron and ion acceleration is markedly influenced by the mass ratio. At low \alf ic Mach numbers ($\Ma \approx 5.3$), the realistic mass ratio allows for efficient electron acceleration through the destabilization of intermediate-scale instabilities, whereas the reduced mass ratio suppresses this process. Conversely, at high Mach numbers ($\Ma \approx 21.3$), the reduced mass ratio leads to an unrealistically high flux of accelerated ions and more efficient electron acceleration compared to the realistic $m_r$, indicating a sensitivity of acceleration mechanisms to $m_r$. The acceleration efficiency of electrons and ions into non-thermal energies further underscores the mass ratio's impact. While ion acceleration efficiency remains approximately 11-12\% across all simulations, electron acceleration efficiency significantly varies. Realistic mass ratio simulations at low Mach numbers show the highest electron acceleration efficiency, whereas at higher Mach numbers, reduced mass ratio simulations show greater efficiency, albeit potentially non-physical.
\end{enumerate}

An important caveat is that the simulations we presented here run up to about \( \sim 4 \times 10^4 \, \omega_i^{-1} \), which is significantly longer than most kinetic simulations we are aware of in the literature. However, when applied to SNR shocks in the ISM with a particle density of 0.1 particles/cm\(^3\), this simulation timescale corresponds to approximately 1.5 minutes—much shorter than the typical age of these shocks, which ranges from \(10^3\) to \(10^4\) years. In other words, we simulate only the first 1.5 minutes of shock evolution. What we observe from these simulations is that, during this initial period, some of these shocks inject electrons into very high energies suitable for the DSA process, with varying efficiencies. However, the ultimate maximum energy remains beyond what we can infer from these simulations.

In conclusion, our findings emphasize the necessity of employing realistic ion-to-electron mass ratios in particle-in-cell simulations to accurately capture dissipation dynamics in collisionless shocks. The disparities observed in thermal and non-thermal energy dissipation between reduced and realistic mass ratios highlight the crucial role of using realistic mass ratios to accurately model shocks in astrophysical environments.

\section*{Author ORCID}
M.\ Shalaby, \href{https://orcid.org/0000-0001-9625-5929}{https://orcid.org/0000-0001-9625-5929}

\section*{Acknowledgments}
We acknowledge the anonymous referee's suggestions, which helped improve the presentation and interpretation of our simulations.
MS acknowledge support by the European Research Council under ERC-AdG grant PICOGAL-101019746.
This work was supported by the North-German Supercomputing Alliance (HLRN) under projects bbp00046 and bbp00072.

{{\small
\bibliographystyle{aasjournal} 
\bibliography{refs}
}}

\end{document}